\documentclass[twocolumn,showpacs,preprintnumbers,amsmath,amssymb]{revtex4}
\usepackage{graphicx}
\usepackage{dcolumn}
\usepackage{bm}
\usepackage{tabularx}
\input{epsf.tex}

\begin{document}

\title{Search for narrow resonances in $\pi p$ elastic scattering from the EPECUR experiment.}
\author{A.~Gridnev,$^{1,*}$ I.G.~Alekseev,$^{2,5}$ V.A.~Andreev,$^{1}$ I.G.~Bordyuzhin,$^{2}$ W.J.~Briscoe,$^{3}$ 
Ye.A.~Filimonov,$^{1}$ 
V.V.~Golubev,$^{2}$  D.V.~Kalinkin,$^{2}$ L.I.~Koroleva,$^{2}$ N.G.~Kozlenko,$^{1}$ V.S.~Kozlov,$^{1}$
 A.G.~Krivshich,$^{1}$ V.A.~Kuznetsov,$^{1}$ B.V.~Morozov,$^{2}$ V.M.~Nesterov,$^{2}$ D.V.~Novinsky,$^{1}$   
 V.V.~Ryltsov,$^{2}$ M.~Sadler,$^{4}$ I.I.~Strakovsky,$^{3}$ A.D.~Sulimov,$^{2}$ 
V.V.~Sumachev,$^{1}$ D.N.~Svirida,$^{2}$ V.I.~Tarakanov,$^{2}$ V.Yu.~Trautman,$^{2}$ R.L.~Workman$^{3}$ 
}
\vspace{0.5 cm}

\affiliation{$^1$Petersburg Nuclear Physics Institute, 188300 Gatchina, Russia}
\affiliation{$^2$Institute for Theoretical and Experimental Physics, Moscow, 117218, Russia}
\affiliation{$^3$The George Washington University, Washington, DC 20052, USA}
\affiliation{$^4$Abilene Christian University, Abilene, Texas, 79699-7963, USA}
\affiliation{$^5$National Research Nuclear University MEPhI, Moscow, 115409, Russia}
\date{\today}

\begin{abstract}

The analysis of high-precision $\pi^{\pm}p \to \pi^{\pm}p$ cross section data from 
the EPECUR Collaboration based on the multichannel  $K$-matrix approach is presented.
The sharp structures seen in these data are studied in terms of both  
opening thresholds and new resonance contributions. Some prominent features are
found to be due to the opening $K\Sigma$ channel. However, 
a complete description of the data is improved with the addition of 
two narrow resonant structures
at $W\sim 1.686$ and $W\sim 1.720$ GeV.  These structures are interpreted 
as manifestations of $S_{11}$ and $P_{11}$ resonances. The underlying nature 
of the observed phenomena is discussed. 

\end{abstract}
\thanks{Electronic address: gridnev@pnpi.spb.ru}
\maketitle

A major challenge in the domain of hadronic physics is the understanding
of states not having the standard $\bar{q} q$ and $qqq$ structures existing in the
traditional Constituent Quark Model (CQM). The prediction
of an antidecuplet of exotic particles (pentaquarks) within 
the framework of the Chiral Soliton Model ($\chi$SM)~\cite{dia} spawned major
experimental efforts worldwide. 
Recently this interest in pentaquarks has been renewed with the claim of 
a charmed pentaquark by the LHCb Collaboration~\cite{lhcb}. It is an open question
whether this newly-discovered state may have its partners at lower masses.

Beginning with the first pentaquark announcement, from LEPS collaboration~\cite{LEPS}, 
there were many reports confirming 
the observation of the lightest member of the proposed antidecuplet, 
the $\Theta ^+$(1538) baryon~\cite{hicks}. In 2004, the Particle Data Group quoted it 
as an established 3-star particle~\cite{pdg2004}. Somewhat later, however, 
most of these results 
were announced to be statistical fluctuations~\cite{hicks}.  Nonetheless, three groups 
LEPS~\cite{leps2}, DIANA~\cite{diana}, and SVD-2~\cite{svd},
still insist on this finding. In 2012, a part of the CLAS 
Collaboration reported a new high-statistics signal which could be associated with
the $\Theta ^+$~\cite{ama}. More recently, however, 
an experiment at J-PARC has found no evidence for this particle~\cite{jparc}.

In 2004, a modified SAID PWA of $\pi N$ scattering data 
allowed for two $P_{11}$ candidates for the second member of the antidecuplet,
the non-strange pentaquark, with masses near $1.68$ and $1.73$ GeV~\cite{arndt}. To be compatible with
the data existing at that time, these candidate 
states were required to be very narrow and have a small branching to $\pi N$. 
In this context, the 
observation of a narrow enhancement at $W \sim 1.68$ GeV in $\eta$ photoproduction on the neutron (the 
so-called ''neutron anomaly") appeared to be an important piece of the puzzle. 
The effect was first observed at GRAAL~\cite{gra} and 
then confirmed by the LNS~\cite{kas}, CBELSA/TAPS~\cite{kru,kru2} and A2@MAMI~\cite{mainz1} Collaborations. 
This structure was not seen in the previous measurements of $\eta$ photoproduction
on the proton~\cite{etap}.  
Recent precise measurements of the cross section
for this reaction, at A2@MAMI-C, have revealed a narrow dip at 
this same energy ~\cite{mainz2}.
A narrow resonance-like structure at $W\sim 1.685$ GeV was also observed
in the $\gamma p \to \eta p$ beam asymmetry data  from GRAAL~\cite{acta}. 
A narrow peak at this energy was found
in Compton scattering on the neutron $\gamma n \to \gamma n$~\cite{comp}
while neither peak was seen in the $\gamma n \to \pi^0 n$ cross section~\cite{pi0}.

This whole assembly of experimental findings has generated a number of explanations.
In line with the pentaquark hypothesis, these may signal a nucleon resonance 
with unusual properties: a mass $M\sim 1.68$~GeV, a narrow ($\Gamma \leq 25$  MeV) 
width, a strong photo-excitation on the neutron, and a suppressed decay 
to $\pi N$ final state~\cite{az,tia,kim,arndt,mart}.
The properties of this putative resonance coincide surprisingly well with those
expected for the second member of the antidecuplet, the non-strange $P_{11}$ pentaquark~\cite{max,dia1}.
However, contradictory explanations also exist, with several groups explaining the bump in 
the $\gamma n \to \eta n$ cross section in terms of i) the interference of 
well-known and broader resonances ~\cite{int} or
ii) the sub-threshold $K\Lambda$ and $K\Sigma$ production (cusp effect)~\cite{dor}. 
Therefore it is of interest to reexamine this problem using elastic $\pi N$
scattering data.

Much of our knowledge of the baryon resonances was obtained by through the analysis of 
$\pi N$ scattering.
In general, theory predicts only weak couplings of pentaquark states
to the elastic $\pi N$ channel. Therefore, experimental data should be of very high precision. 
On the other hand the analysis of such data would have some advantages:
i) the structure of $\pi N$ amplitude is essentially simpler than that of photoproduction;
ii) the $\pi N$ partial waves are quite well known from phase shift analysis;
iii) there is isospin symmetry in the $\pi N$ system.

In the years from 2005 to 2013, the EPECUR Collaboration measured $\pi^{\pm}p \to \pi^{\pm}p$  elastic scattering
over an energy range of $ p_{lab} = 800 - 1300$ MeV/c and for angles $\theta_{cm}$ from 40 to 120
degrees~\cite{epe1}.
In total, about $10000$ new data points have been
obtained. These data have been produced with a momentum resolution of $\sim 1$ MeV and with
$\sim 1\%$  statistical errors. 

The $\pi^{-}p \to \pi^{-}p$ data revealed two narrow structures, at  $W\sim 1.686$ and
at $W\sim 1.720$ GeV,  which were not seen in $\pi^+ p$ scattering~\cite{grid}. This clearly shows that 
the observed structures appear in the isospin I=1/2 sector only.
It is interesting to note that a structure at $W\sim 1.720$ GeV was also recently 
found in Compton scattering off the proton~\cite{comp1}
and $\eta$-photoproduction off the neutron~\cite{kru1}.

In Ref.~\cite{grid}, a preliminary analysis of the data from Ref.~\cite{epe1} was presented, with the 
finding that these
structures could be described by two narrow (width $\sim$ 25 MeV)
$S_{11}$ and $P_{11}$ resonances. 
In this paper,  an analysis of the full EPECUR database ~\cite{epe1} is presented.
Here we attempt to explain observed structures in terms of both couplings to inelastic channels 
and resonance contributions. 
For that purpose, we employ a K-matrix approach based on the effective Lagrangians
described in Refs.~\cite{goud,grko}, and applied to both 
$\pi N$ scattering and photoproduction in Ref.~\cite{feus}.

It is assumed that the K-matrix, as a solution to equations yielding the scattering amplitude,
can be described in terms of a sum of the tree-level Feynman diagrams with vertices obtained
from an effective Lagrangian.
The model includes four-star PDG ~\cite{PDG14} resonances in the $s$- and $u$-channels and
$\sigma$, $\rho$, $a_{0}$ and $K^*$ exchange in the $t$ channel. To describe the high energy tail in $\pi^+ p$ data,
the three-star P33(1900) resonance was also included.  
Two new isospin-1/2 resonances were added, as well, to reproduce observed structures in the $\pi^- p$ data,
as we describe below. In total, 
the 5-channel analysis took into account elastic,
$2 \pi $ (effective),  $\eta n$ , $K \Lambda $, and $K \Sigma $ production. 

As the main goal of this work was to explore the nature of narrow structures in $\pi^- p$ elastic 
scattering, a detailed description of inelastic channels 
was not attempted. This reduced the number of free parameters, resonance masses and couplings, used
in the fits.
The employed database included the EPECUR data, the total 
cross-sections for $\pi^- p \to \eta n$ ~\cite{exeta}, and data for the differential cross sections
of $\pi^- p \to  K\Lambda $ and $\pi^- p \to  K\Sigma $ ~\cite{exkl}. 
To achieve the consistency with the data on elastic $\pi N$  scattering, at the energies below the EPECUR
data, single-energy solutions from the $XP15$ ~\cite{epe1} partial wave analysis (PWA) 
were added to the data base.

\begin{figure}[!ht]
  \centering
    \includegraphics[width=0.5\textwidth]{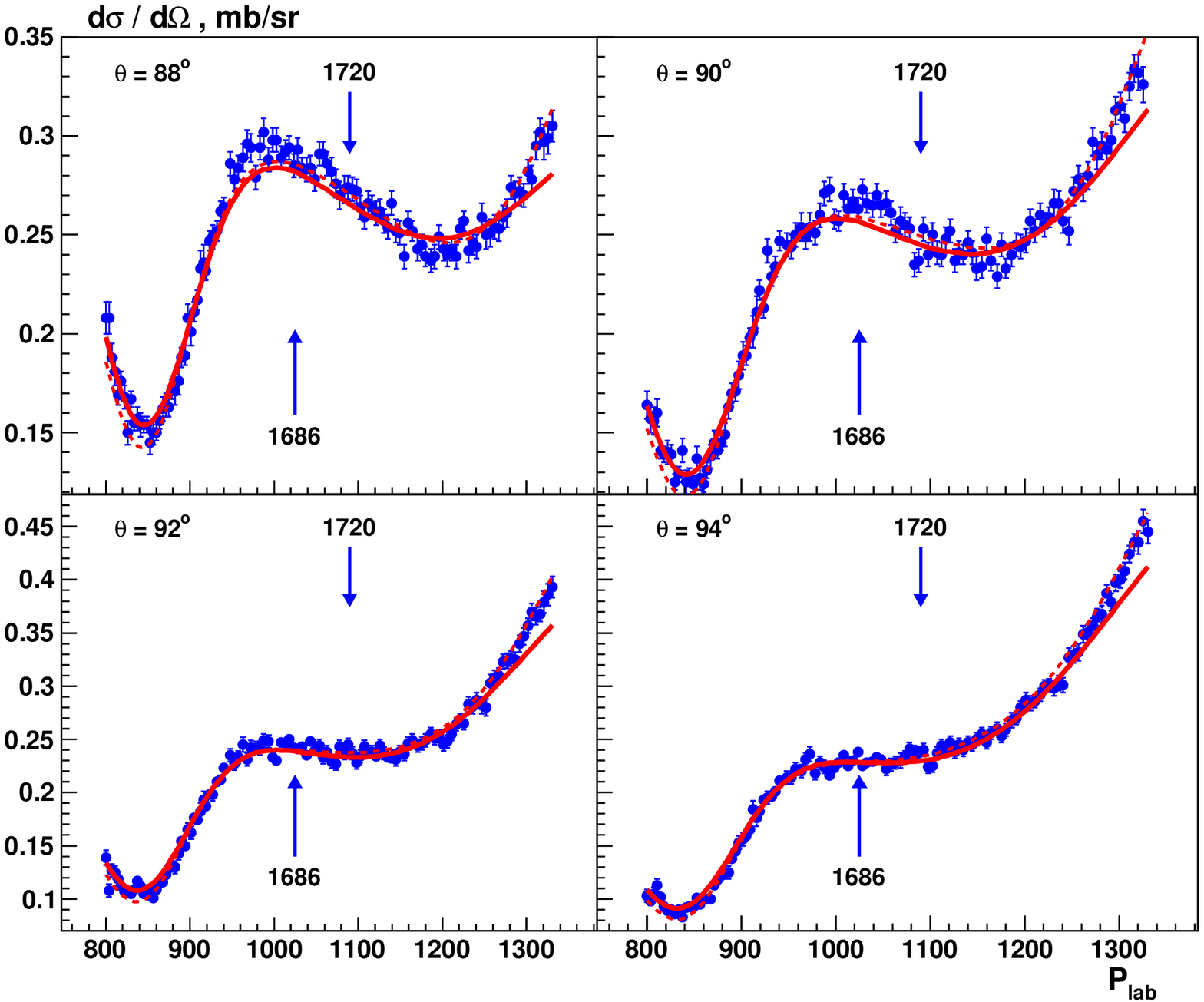}
  \caption{$\pi^+ p$ elastic scattering. Solid lines correspond to the present calculations. The dotted
  lines indicate the XP15 solution.}
  \label{fig:pipp_elast}
  \centering
    \includegraphics[width=0.5\textwidth]{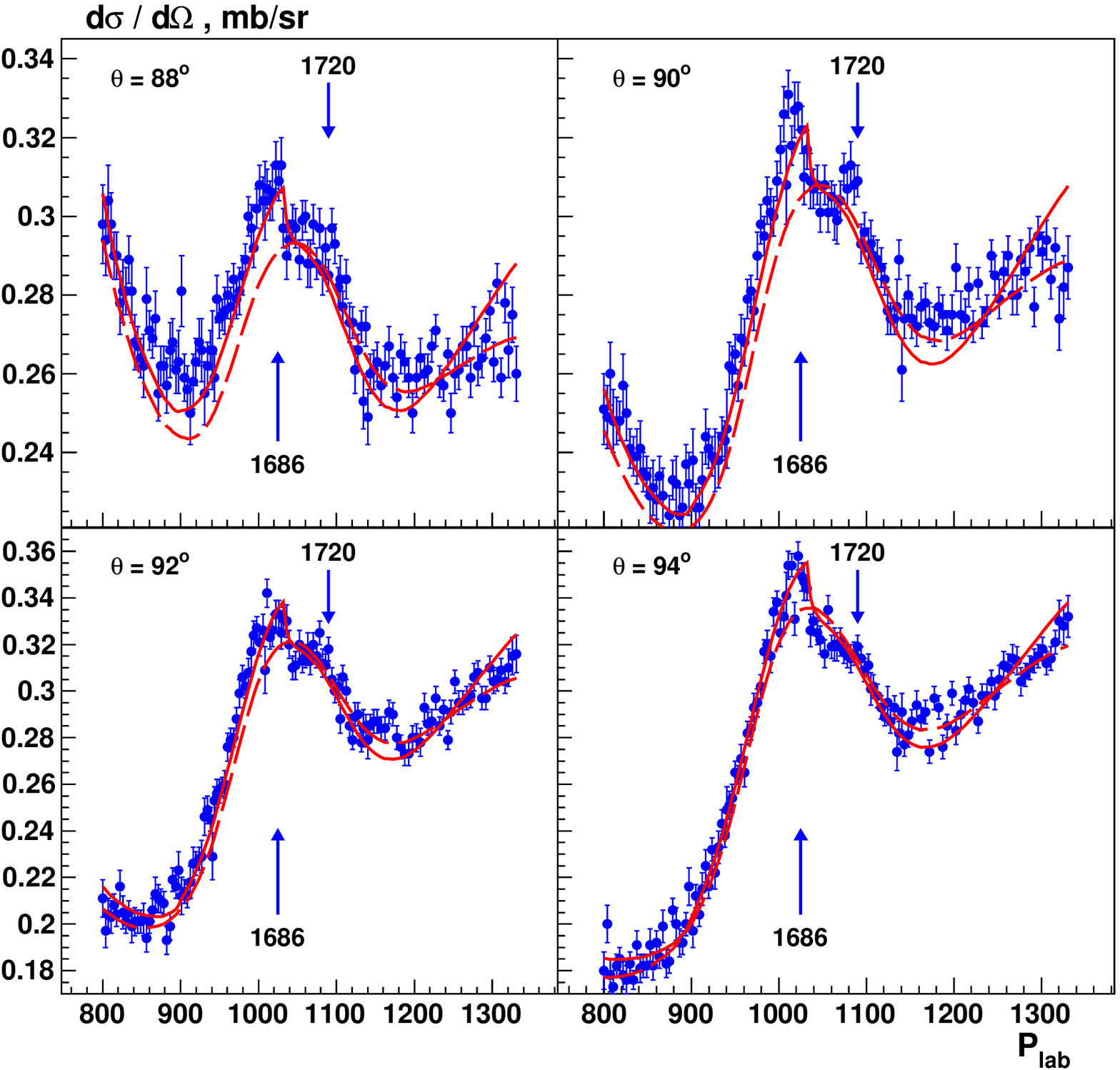}
  \caption{$\pi^- p$ elastic scattering. Solid lines correspond to the present calculations. The dotted
  lines are the XP15 solution.}
  \label{fig:pimp_elast}
\end{figure}

The $XP15$ solution was the result of a SAID PWA analysis which included the EPECUR data.
This solution provided a rather good description of the whole data set getting a
$\chi^2 \sim  3 $ per point. However, a description the abovementioned sharp structures
was absent. This is clear from Figs.~\ref{fig:pimp_elast},
in which the dotted lines correspond to the $XP15$ solution. The results of our calculations
without any narrow resonances are shown
in this figure by the solid lines. It should be noted that the XP15 parameterization included
the inelastic channels $\pi \Delta$, $\rho N$, and $\eta N$, but no $K\Lambda$ or $K\Sigma$
channels. 

Without the inclusion of narrow resonances, the solid lines in  
Fig.~\ref{fig:pimp_elast} reproduce the rapid variation of 
the energy dependence seen in the $\pi^- p$ differential cross section close 
to the $\pi^- p \to  K\Sigma $ threshold at the angles $\sim 90$ degrees.
Such an effect is not seen in the $\pi^+ p$ data.

\begin{figure}[!ht]
  \centering
    \includegraphics[width=0.5\textwidth]{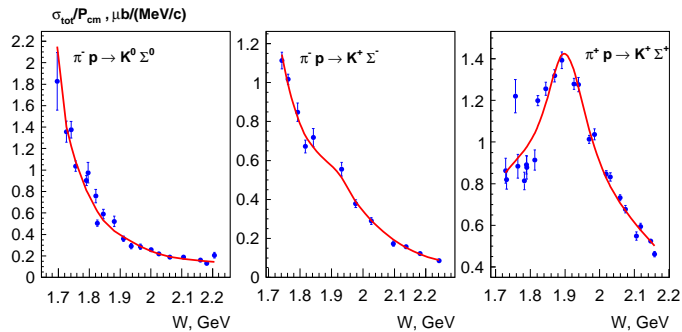}
  \caption{Total cross section for $\pi p \to K \Sigma$. The data are from ~\protect\cite{exkl}.
  Solid lines are from the present work.}
  \label{fig:KSigma_tot}
\end{figure}

A qualitative explanation of this phenomenon is evident from 
Fig.~\ref{fig:KSigma_tot}, in which the energy dependence of the total $\pi p \to  K\Sigma$ cross section
for different charged states is shown.
One can see that the $\pi^- p \to  K^0\Sigma^0$ and $\pi^- p \to  K^+\Sigma^-$ plots vary 
rapidly near the threshold 
$W\sim 1690$ GeV, while the energy dependence of the $\pi^+ p \to  K^+\Sigma^+$ reaction is more smooth,
and therefore does not generate sharp structures in the $\pi^+ p$ scattering data.
Our results for the $\pi p \to  K\Sigma$ total cross section are shown in Fig.~\ref{fig:KSigma_tot}
by solid lines. The present calculation reproduces these data quite well.

\begin{figure}[!ht]
  \centering
    \includegraphics[width=0.5\textwidth]{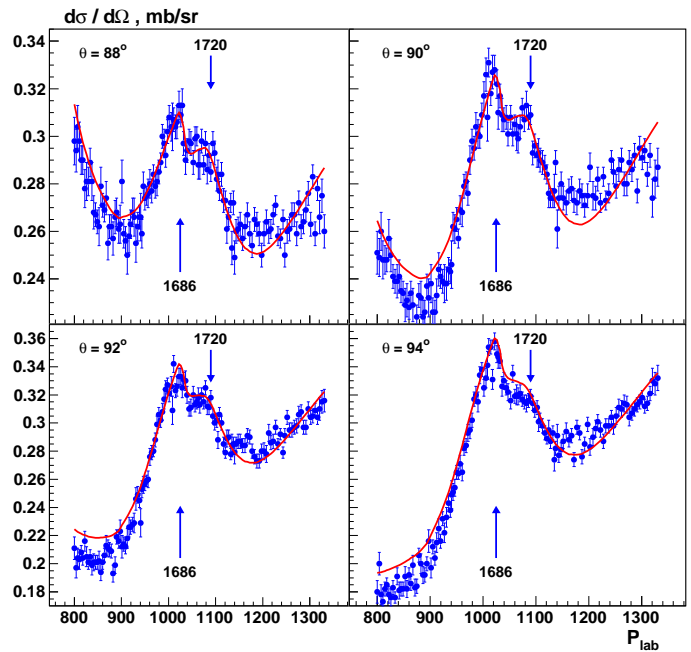}
  \caption{$\pi^- p$ elastic scattering with added resonances. Solid line gives the present calculation.
  }
  \label{fig:pimp_res}
\end{figure}

   As a next step, two resonances were added in an attempt to improve the fit quality around $90^{\circ}$. 
Here, the overall $\chi^2$ per datum is not a good parameter to estimate the quality of the fit, as the 
structure is evident in only $\sim 200$ data points among $5000$ in total. Thus, the overall $\chi^2$ would
be overwhelmed by the quality of fit to the background behavior.
To compare the different fits with additional 
resonances, $\chi^2$ in the restricted energy interval of $ p_{lab}$ = 980 - 1140 MeV/c was calculated.
While different quantum numbers for the added resonances were tested, only S11 for the first and P11 for the 
second gave a reasonable $\chi^2$. The inclusion of these resonances lead to a significant
improvement of $\chi^2 \sim 1.5 $ as compared with 
$\chi^2 \sim  2.6$ for the background. 
The results are shown in Fig.\ref{fig:pimp_res} and Table 1. 

\begin{table}[!ht] \footnotesize
  \centering
	\caption{Resonance parameters.}
	\label{tab:par_res}

	\begin{tabular}{|p{1.5cm}|c|c|}
		\hline
		& \hspace{0.5cm}S11 ~~~~~~&\hspace{0.5cm}P11 ~~~~~~~\\	
		& (MeV)	&  (MeV)\\	\hline
	Mass	& 1688	& 1724	\\	\hline
$\Gamma_{el}$	& 5.0	& 8.5	\\	\hline
$\Gamma_{\eta n}$& 2.3	& 19.8	\\	\hline
$\Gamma_{2\pi}$	& 0.3	& 7.1	\\	\hline
$\Gamma_{K\Lambda}$& 10.0 & 4.8	\\	\hline
$\Gamma_{K\Sigma}$& --	& 4.0	\\	\hline
$\Gamma_{tot}$	& 17.6	& 44.2	\\	\hline
	\end{tabular}
\end{table}

Both resonances have the small widths and the small couplings to the elastic
 $\pi N$ channel. This is in agreement to the predicted properties of the non-strange pentaquark state,
the second member of the antidecuplet.

\begin{figure}[!ht]
  \centering
    \includegraphics[width=0.5\textwidth]{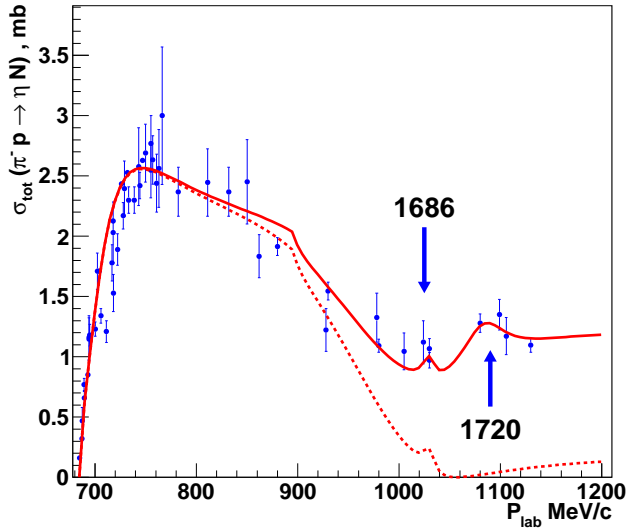}
  \caption{Comparison of the measured and calculated $\pi^- p \to \eta n$ total cross sections. 
The solid line presents the calculations from this work.
The dotted line indicates the S-wave contribution.}
  \label{fig:etan_tot}
\end{figure}

Having concentrated on structure in $\pi N$ scattering, it is important to see how the added resonances would  
appear in inelastic channels. In 
Fig.~\ref{fig:etan_tot}, the total cross section for $\pi^- p \to  \eta n $ is presented.

One may see that the data are not in conflict with resonance contributions but also do not prove their existence.
The dotted line in Fig.~\ref{fig:etan_tot} gives
the S-wave contribution to $\pi^- p \to \eta n$. 
As was shown in Ref.~\cite{grko}, a minimum of the S-wave contribution near 
 $P_{lab} \sim 1050$ MeV/c could be explained through interference of the $S_{11}(1535)$ and 
$S_{11}(1650)$ resonances. Different signs for the coupling constants of the $\eta $ meson with 
these resonances was found, in agreement with Refs.~\cite{int}. 
But opposite to these works, the interference does not produce any sharp peak 
in the $\pi^- p \to  \eta n $ reaction. Moreover, a very small ($1\% $) branching ratio of the S11(1650) resonance 
to $ \eta n $ is found in the present work.
  
\begin{figure}[!ht]
  \centering
    \includegraphics[width=0.5\textwidth]{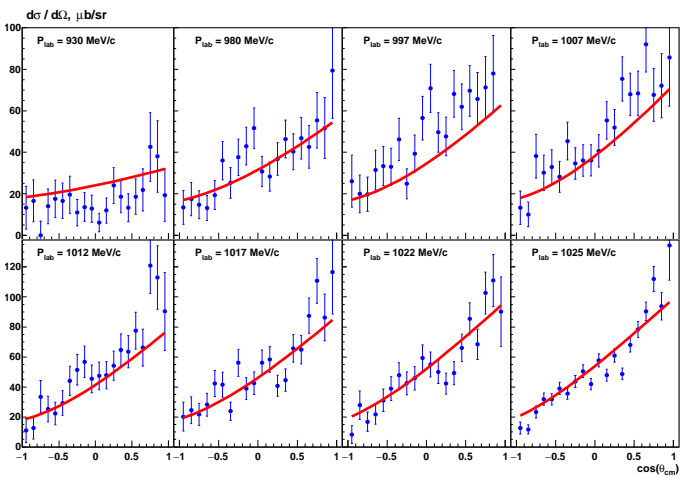}
  \caption{$\pi^- p \to K \Lambda$ differential cross section. Solid line - present calculations.}
  \label{fig:KLambda_diff}
\end{figure}

\begin{figure}[!ht]
  \centering
    \includegraphics[width=0.5\textwidth]{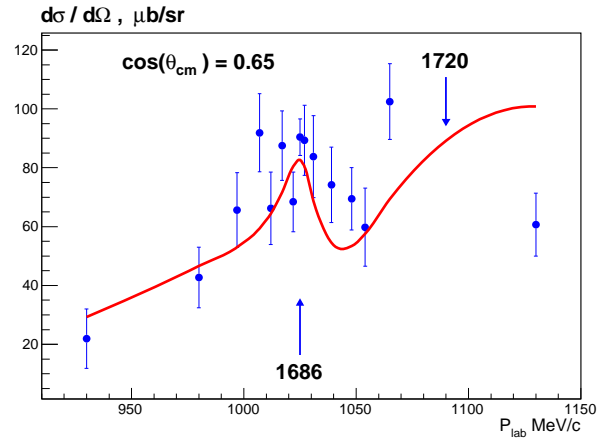}
  \caption{Energy dependence of $\pi^- p \to K \Lambda$ differential cross section. Solid line - present calculations.}
  \label{fig:KLambda_plab}
\end{figure}

Figs.~\ref{fig:KLambda_diff} and ~\ref{fig:KLambda_plab} show the results for  $\pi^- p \to  K \Lambda $ 
differential cross sections and their energy dependence for cos($\theta_{cm}$)=0.65. Again, the existing large 
experimental errors do not make a definite conclusion possible 
regarding the existence of the added narrow resonances.


We conclude that two narrow structures observed in elastic $\pi^- p$ scattering can be explained
by a combination of threshold effects and 
two narrow resonances S11(1686) and P11(1720). These contributions we discuss separately below.

Concerning the narrow resonance contributions,
narrow structures are also seen
in Compton scattering ~\cite{comp} and $\eta$ photoproduction off the neutron~\cite{mainz1}. 
What is the nature of these structures? The interference of well-known 
resonances suggests a delicate relation 
between incoming and outgoing vertexes. 
It is unlikely that this relation is valid for all three reactions, namely $\pi^- p$ scattering,
Compton scattering and  $\eta$ photoproduction. Further work is required before a
definitive conclusion can be drawn concerning this possibility.

Another contribution is available via the cusp effect,  i.e. the 
influence closed channels on incoming amplitude due to the  analyticity condition. 
This element requires further development and remains a hypothesis which requires further detailed verification.
It is worth noting that not all threshold effects result in sizable cusp effects. 
For instance, no clear structure is seen in $\pi^- p $ elastic scattering near 
the $K \Lambda $ threshold ($P_{lab}$=900 MeV/c).

Finally, in the energy region around 1686 MeV, possible electromagnetic effects must be 
taken into the consideration. Indeed, just below the K$\Sigma$ threshold, a bound  
atomic-like state of  $K^+\Sigma^-$ could be created. If it exists, this state could in fact 
be seen in the $\pi^- p$ and $\gamma n$ reactions
only. The existence of these electromagnetic effects could be checked, for example,
by measuring the cross-sections for two isospin-symmetric reactions $\pi^- p \to  \eta n $
and $\pi^+ n \to  \eta p $. Accordingly, the isospin symmetry 
cross-sections of reactions should be the same but the $K^+\Sigma^-$ system would exist 
for the first reaction only. 
No such effect exists for a narrow $P_{11}(1724)$, and we consider this resonance, which has the 
nucleon quantum numbers to be the best candidate for the non-strange member of an exotic antidecuplet. 
New high precision experimental data on $\pi^- p \to  K\Lambda $ and $\pi^- p \to  \eta n $ are needed to 
achieve a decisive conclusion.

 This work was partially supported by Russian Fund for
Basic Research grants 12-02-00981a,9-02-00998a, 05-02-17005a
and by the U.S. Department of Energy Grant DE-SC0014133.


\end{document}